\documentclass[useAMS,usenatbib,usegraphicx,referee]{mn2e}
\usepackage{graphicx,amssymb,amsmath,times,verbatim}
\usepackage{color}
\usepackage[T1]{fontenc}

\title[]{High Energy Emission Processes in OJ 287 during 2009 Flare}
\author[]{Pankaj Kushwaha$^{1}$\thanks{E-mail:pankaj563@tifr.res.in}, S. Sahayanathan$^{2}$, K. P. Singh$^{1}$\\
$^{1}$Department of Astronomy and Astrophysics, Tata Institute of Fundamental Research, Homi Bhabha Road, 
Mumbai 400005, INDIA\\$^{2}$ Astrophysical Sciences Division, Bhabha Atomic Research Centre , Mumbai 400085, INDIA}

\begin{document}
\maketitle

\begin{abstract}
The broadband spectrum of a BL Lac object, OJ 287, from radio to $\gamma$-rays obtained during a major $\gamma$-ray 
flare detected by \emph{Fermi} in 2009 are studied to understand the high energy emission mechanism during this episode.
Using a simple one-zone leptonic model, incorporating synchrotron and inverse Compton emission processes,
we show that the explanation of high energy emission from X-rays to $\gamma$-rays, by considering a
single emission mechanism, namely, synchrotron self-Compton (SSC) or external Compton (EC) requires unlikely 
physical conditions. However, a combination of both SSC and EC mechanisms can reproduce the observed high energy spectrum satisfactorily. 
Using these emission mechanisms we extract the physical parameters governing the source and its environment. Our study 
suggests that the emission region of OJ 287 is surrounded 
by a warm infrared (IR)  emitting region of $\sim 250 \, K$. Assuming this region as a spherical cloud illuminated by an accretion 
disk, we obtain the location of the emission region to be $\sim 9~ pc$. This supports the claim that the 
$\gamma$-ray emission from OJ 287 during the 2009 flare arises from a location far away from the central 
engine as deduced from millimeter-gamma ray correlation study and very long baseline array images.

\end{abstract}


\section{INTRODUCTION}
\label{sec:intro}
BL Lacs are a class of radio loud active galactic nuclei (AGN) with no/weak emission line features \citep{1995PASP..107..803U}. 
They are classified along with flat spectrum radio quasars (FSRQ) as blazars.
BL Lacs are characterised by a non-thermal spectra extending from radio to $\gamma$-rays with many of them detected 
even up to GeV/TeV energies \citep{2011arXiv1110.0040W} \footnote{http://tevcat.uchicago.edu/}. Their spectral energy 
distribution (SED) is bimodal with a low energy peak in IR-X ray wavelength and a high energy one  
at $\gamma$-rays. Based on the location of low energy peak they are classified as low energy peaked BL Lacs (LBL), intermediate 
energy peaked BL Lacs (IBL) and high energy peaked BL Lacs (HBL) \citep{1995ApJ...444..567P, 1998MNRAS.299..433F}. The observed short time 
variability of the order of days to minutes and detection of very high energy $\gamma$-rays demand
the emission to arise from a relativistic jet close to the line of sight of the observer \citep{1995MNRAS.273..583D}.
Moreover, the constraints obtained from the variability timescale suggest,
the emission region to be located  at sub-parsec scales from the central engine. 
The strong polarization observed in radio and optical bands, and the non-thermal 
nature of the spectrum indicate that the radio-to-X-ray emission is of synchrotron origin 
due to cooling of a power-law distribution of electrons in a magnetic field. The higher energy emission
is then generally attributed to synchrotron self-Compton (SSC) emission where the population of
electrons responsible for synchrotron emission will further scatter off the synchrotron photons to higher energies 
by inverse Compton process. However, for certain 
BL Lacs one needs to consider the inverse Compton scattering of photons external to the jet in 
order to explain the high energy emission \citep{2011ApJ...726...43A, 2012ApJ...751..159A}. Besides these models, 
there also exist other models where the high energy emission 
is believed to be the result of hadronic processes \citep{1998Sci...279..684M,2003APh....18..593M,2009ApJ...703.1168B}.

The relativistic jets of a few BL Lacs are resolved in the high resolution radio maps and 
often show knot like features \citep{2004ApJ...600..127G, 2011ApJ...729...26M}. 
In many misaligned AGN jets, Fanaroff and Riley type I and II (FR I and FR II) \citep{1974MNRAS.167P..31F}, such knots are even observed in optical and X-ray maps at kilo
parsec scales \citep{2002ApJ...571..206S,2001ApJ...556L..79P}. Often the location of these optical/X-ray knots are coincident with the ones seen in radio.
For these knots, the radio-to-optical emission 
is generally attributed to synchrotron emission whereas X-ray emission can be an extension of synchrotron radiation itself or arise due to inverse Compton 
scattering of soft target photons \citep{2000ApJ...544L..23T}. If X-ray flux is above the extrapolation of radio to optical
flux then X-ray emission is explained through inverse Compton process else synchrotron emission model is 
accepted \citep{2002ApJ...571..206S}. For the case when X-ray emission is due to inverse Compton 
processes, SSC interpretation requires large jet power and magnetic field lower than the equipartition
value \citep{2000ApJ...542..655C}. On the other hand inverse Compton scattering of external photons may be a viable option.
At kilo parsec scales the plausible external photon field for the inverse Compton scattering can be  the
Cosmic microwave background radiation (CMBR). Explanation of X-ray emission through 
inverse Compton scattering of CMBR (IC/CMBR) requires less jet power and near equipartition magnetic field \citep{2000ApJ...544L..23T} and this process 
is widely accepted though it faces various criticisms \citep{2004ApJ...613..151A,2006ARA&A..44..463H}. 
At parsec scales, the high energy emission 
can be due to inverse Compton scattering of radiation from the nuclear region and/or starlight \citep{2006MNRAS.370..981S}.
When X-ray emission is due to synchrotron
process, the underlying particle distribution requires a broken power-law in order to explain
the observed spectrum \citep{2002ApJ...568..133W,2007ApJ...668L..23L}. 
A broken power-law particle distribution can be formed by a continuous injection of plasma into a 
cooling region or through multiple acceleration processes \citep{2008MNRAS.388L..49S,2003ApJ...588L..77S}.

OJ 287 (z = 0.306) is one of the well studied BL Lac object (LBL) with a peculiar  
periodic outbursts at an interval of roughly 12 years \citep{1996A&A...305L..17S}.
This behaviour suggested the possible presence of a binary super massive black hole (SMBH) system at the nucleus of OJ 287.
\citet{1988ApJ...325..628S} explained these outbursts as a result of tidal disturbances in the accretion disc of the  
primary black hole caused by the secondary. Later \citet{1996A&A...315L..13S} performed a detailed study of the optical light curves 
during these periodic episodes and found that the outbursts are double-peaked.  They explained this feature as a result of the double impact of the secondary 
black hole on the accretion disk of the primary while orbiting around the latter in the binary black hole system \citep{1996ApJ...460..207L}.
This model was later modified to accommodate new data obtained during 2005-2007 outburst \citep{2007ApJ...659.1074V,2006ApJ...646...36V,2009ApJ...698..781V,2011ApJ...729...33V}.

Recently OJ 287 was observed extensively through several campaigns during 2005-2010 (\cite{2011AcPol..51f..76V} and references therein). The observations were
primarily oriented towards the detection of the outburst around, confirming the prediction of binary black hole
system originally proposed by \cite{1988ApJ...325..628S}. \cite{2011ApJ...726L..13A} studied the $\gamma$-ray flare of OJ 287 
during 2008-2010 along with observations in other energy bands around the same period. They found a strong correlation
between $\gamma$-ray and the millimeter emission during the two major $\gamma$-ray flares. 
Further the VLBA (Very Long Baseline Array) study suggested that the millimeter flares are associated with the ejection of superluminal patterns
from a stationary knot C1 \citep{2011ApJ...726L..13A}. Based on these facts they argued that the location of the $\gamma$-ray 
emission is linked with the knot C1. From the separation between knot C1 and an inner knot C0, located
close to the nucleus, they concluded that the $\gamma$-ray emission region must be at a distance $>14~ pc$
from the central engine. At such a distance, $\gamma$-ray emission mechanism can be due to either SSC 
or EC scattering of IR photons from a dusty torus (EC/IR).

In the present work we have studied the plausible high energy (X-ray and $\gamma$-ray) emission mechanisms in OJ 287 
during a major flare in 2009.
As suggested by \citet{2011ApJ...726L..13A}, this emission may also be associated with the ejection of superluminal patterns from the 
knot C1. We divided the flare light curve during this period into three different states and obtained their average flux
in various energy bands. A simple emission model was then used to study the observed broadband spectrum 
corresponding to different states. We have exploited the available information obtained through simultaneous/contemporaneous 
multi wavelength observations which are sufficient to 
extract the physical parameters of the source. We analysed the possibility of reproducing the broadband SED of OJ 287  
considering different combinations of emission mechanisms like 
i) synchrotron and SSC, ii) synchrotron and EC/IR, and iii) synchrotron, SSC and EC/IR 
and present our results here. Below we first describe the data analysis 
technique (\S \ref{sec:obser}). In section \S \ref{sec:emissmech} we study the processes responsible for the high energy emission mechanism. We 
discuss the results obtained in \S \ref{sec:discussion}. A flat $\Lambda CDM$-cosmology 
with $\Omega_m$ = 0.3, $\Omega_\Lambda$ = 0.7 and $H_0$ = 71 ~$km ~s^{-1}Mpc^{-1}$ is assumed throughout 
the paper.

\section{DATA ANALYSIS}
\label{sec:obser}
We have used the publicly available multi-wavelength data from \emph{Fermi-LAT} (Large Area Telescope) and \emph{Swift-XRT} (X-Ray Telescope)
along with optical and radio data from various blazar monitoring programs during the flaring episode (MJD: 55110-55185) of OJ 287 
in 2009.

\subsection{Gamma-ray data}
The  $LAT$ on the \emph{ Fermi Gamma-ray Space Telescope} is a pair production telescope sensitive to $\gamma$-rays energies 
from 30 MeV to $>$ 300 GeV \citep[]{2009b}. Its periodic $\sim$ 3 hours (2 orbit) scan of the entire sky makes it the best instrument to monitor the evolution 
of GeV sources as well as any high energy (HE) phenomenon down to the scanning time scale  thereby helping to understand and 
constrain the HE physics and associated emission processes.

 LAT data of OJ 287 obtained during 2009 flare (MJD: 55110-55185) were analyzed using \emph{Fermi Science} tool version 
 v9r23p1, latest publicly available release during the time of data analysis. Only ``source class events'' (evclass 2) having energy above 100 MeV from 
photon data were considered with the recommended time interval\footnote{http://fermi.gsfc.nasa.gov/ssc/data/analysis/documentation/
Cicerone/Cicerone\_Data\_Exploration/Data\_preparation.html} to make sure that the spacecraft was in normal science data acquisition mode, avoiding 
Earth's limb, South Atlantic Anamoly (SSA) and pointed observations. \emph{Unbinned ~maximum~ likelihood ~analysis} \citep[]{1996ApJ...461..396M} method 
was used to model the photons from a region of interest (ROI) of  $15^\circ$ centered on 
the location of OJ 287 to reconstruct the source energy spectrum. Effects of time selection, energy cut, variation of LAT area with azimuth angle, and 
point-spread function (PSF) corrections were accounted for while generating the exposure map from an annular region of $10^\circ$ around ROI. Sources in the 
region were modeled using LAT second catalog \citep[]{2012ApJS..199...31N} \footnote{OJ 287 is fitted with a simple power-law model}. Pass 7 instrument 
response function with galactic diffuse emission model (gal\_2yearp7v6\_v0.fits) and isotropic background model (gal\_2yearp7v6\_v0.fits), provided 
by the LAT science team were used to model the source spectrum (0.1-300 GeV) keeping integral flux and photon index as free parameters. Source fluxes 
in different energy ranges (100-300MeV, 300MeV-1GeV, 1-3GeV, 3-10GeV) were then extracted by freezing the photon index to the best fit value obtained by the analysis of 0.1-300 GeV data.
Finally, time averaged SED data points were extracted by combining LAT data over the mentioned period (see \S \ref{sec:MWS}) following the procedure as described above.

\subsection{X-ray data}
The \emph{XRT} \citep[]{2005SSRv..120..165B} onboard {\it Swift} is a grazing incidence Wolter type 1 focusing X-ray telescope sensitive  to soft X-ray energies
(0.3-10 keV) . We have used the Photon Counting (PC) pointed data with normal clocking and default window configuration for this study.
 
Event files obtained from \emph{Swift-XRT} database were calibrated and cleaned with standard filtering criteria using \emph{xrtpipeline} (SWXRTDAS version 2.8.0) task and latest calibration files from
\emph{Swift CALDB}. Source photons for spectral analysis were extracted using a circular region of 20 pixel ($\sim 47'', 90\%~PSF$ $at ~1.5~keV$) 
\citep[]{2005SPIE.5898..360M} centered on the source, and background photons from multiple uncontaminated regions around the source. CCD defects and PSF 
corrections were applied using auxillary response file (ARF) generated from \emph{xrtmkarf} task.
The data in the energy bins of the resultant spectrum file (0.5-10 keV) were re-binned using \emph{GRPPHA} with a minimum 5 $\sigma$  significance (statistical only) and 
fitted with a power law model modified by an absorber (phabs) within \emph{XSPEC} (version 12.7.1) by freezing the neutral hydrogen column density ($N_H$) to its 
Galactic value of $2.38\times 10^{20}~cm^{-2}$\citep{2005A&A...440..775K} in the direction of OJ 287. For SED analysis, individual XRT event files during the considered period (see \S \ref{sec:MWS})
were combined using the \emph{XSELECT} task and an average spectrum was extracted. The extracted SED data points were then corrected for Galactic absorption.

\subsection{Optical and Radio data}
Contemporaneous optical and radio data used in this study were taken from archives of various multi-wavelength programs supporting \emph{Fermi} observatory. 
The optical data include V-band photometric data from Arizona-Steward\footnote{http://james.as.arizona.edu/~psmith/Fermi} and near-IR-optical photometric data from 
Yale-SMARTS\footnote{http://www.astro.yale.edu/smarts/glast} (Small and  Medium Aperture Research Telescope System) project. The radio data at 15 GHz and 
43 GHz were obtained from Caltech-OVRO\footnote{http://www.astro.caltech.edu/ovroblazars} (Owens Valley Radio Observatory) and Boston-VLBA 
\footnote{http://www.bu.edu/blazars/VLBAproject.html} project respectively.

The details of data selection and analysis procedure for Yale-SMARTS, Arizona-Steward, Caltech-OVRO and Boston-VLBA data are described in 
\citet[]{2012ApJ...756...13B} (and references therein), \citet[]{2009arXiv0912.3621S}, \citet[]{2011ApJS..194...29R} and \citet[]{2005AJ....130.1418J} respectively.

\subsection{Multi-Wavelength SEDs}\label{sec:MWS}
\begin{figure} 
	\centering
\includegraphics[scale = 0.70]{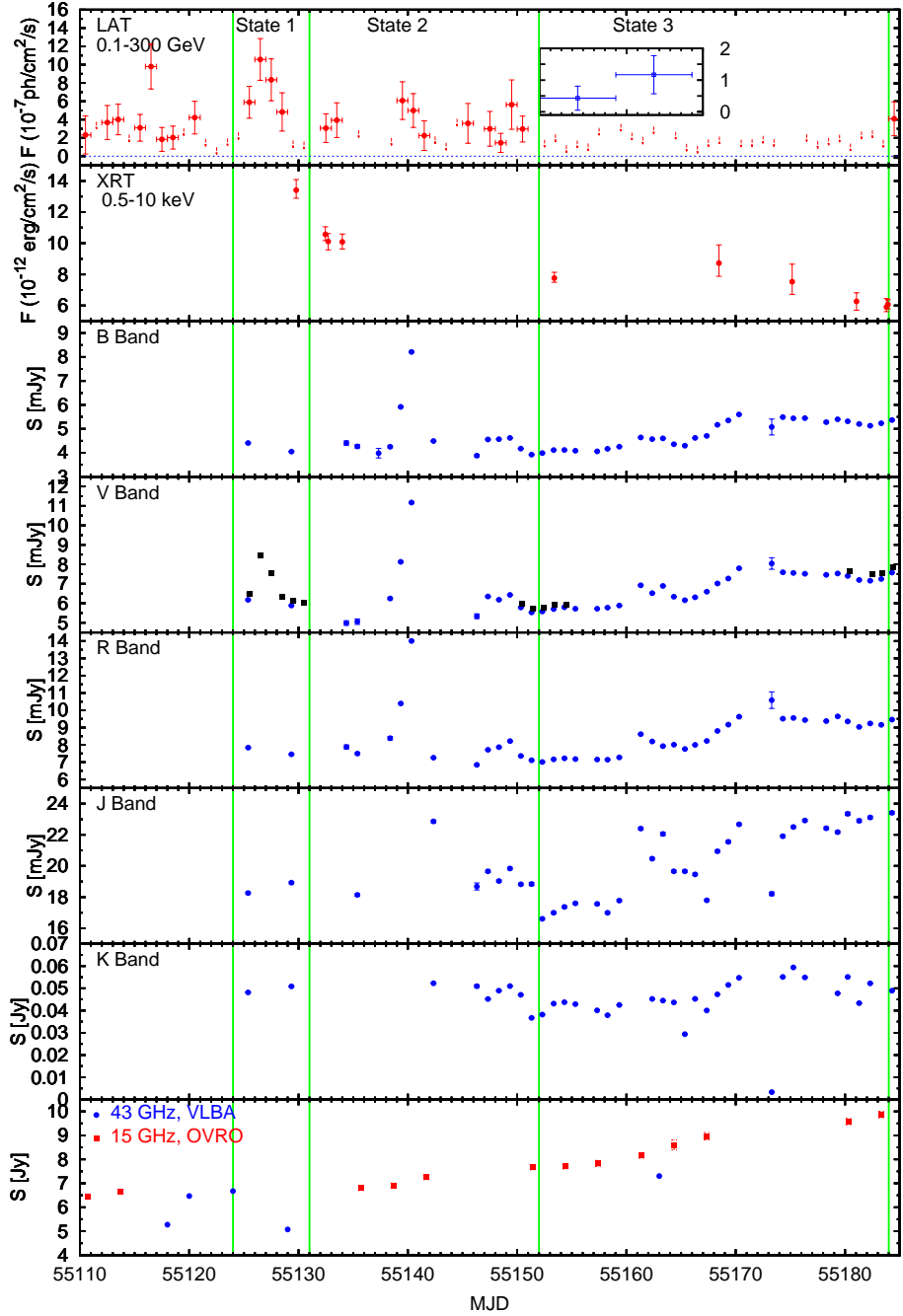}
\caption
{Multi-wavelength light curves of OJ 287 during a flare in 2009 (October 6th-December 20th) from radio to $\gamma$-ray frequencies. The vertical lines delineate the three different states of OJ 287.
 The X-ray (Swift-XRT) and $\gamma$-ray (Fermi-LAT) data are from their respective database. The IR-optical data are reproduced from Yale-SMARTS (filled circles) along with optical V-band data 
from Arizona-Steward (filled squares) \emph{Fermi} follow-up programs. Radio data were obtained from Boston and OVRO blazars' monitoring program as labeled in the figure. The 7 day binned 
LAT data from MJD: 55152-55166 is shown in the inset with the same x-scale.} 
\label{fig:lc}
\end{figure}

Figure \ref{fig:lc} shows the multi-wavelength light curves of OJ 287 during the flare in 2009 as observed by various satellites and ground based observatories (mentioned above). The daily binned 
\emph{LAT} light curve (top panel) corresponds to a detection criteria of  3$\sigma$ (TS $>~ 9$) \citep[]{1996ApJ...461..396M} followed by X-ray, IR, optical and  radio light curves. The inset of Figure
\ref{fig:lc} shows the 7 day binned LAT photon flux with TS $>~9$ for MJD: 55152-55166.

The near correlated variation in different energy bands (visible in the \emph{LAT} and the optical V-band data around MJD 55126 with a hint in X-rays as well) suggests a 
co-spatial origin of radiation 
emphasizing that a single electron population may be responsible for emission throughout the electromagnetic spectrum i.e., from mm (below this frequency 
different regions are believed to be contributing to the radio fluxes \citep{1994ApJ...435L..91M}) to $\gamma$-ray energies. 
Based on the observed $\gamma$-ray variability, nearly correlated variation across different energy
bands and the observation that X-rays follow roughly the same behavior as the $\gamma$-rays, we divided the data in three activity phases: \emph{State 1} or an active phase (MJD: 55124-55131), 
\emph{State 2} or a moderately active phase (MJD: 55131-55152) and \emph{State 3} or a quiescent phase (MJD: 55152-55184). The vertical lines in figure 1 delineate these three different phases of 
the source and the corresponding SEDs are shown in figure 2. The NIR-optical magnitudes were de-reddened following the method given by \citet[]{2011ApJ...737..103S} with $R_V = 3.1$ and 
$E(B-V) = 0.0280 \pm 0.0008$ \citep[]{1998ApJ...500..525S}. Magnitude to flux density conversion was done using the zero-magnitude flux densities given by \citet[]{1998A&A...333..231B}.
The properties of the source and the high energy emission mechanisms are then studied through detailed modelling of these SEDs as described in the next section.

\begin{figure} 
\begin{center}
\includegraphics[scale = 1]{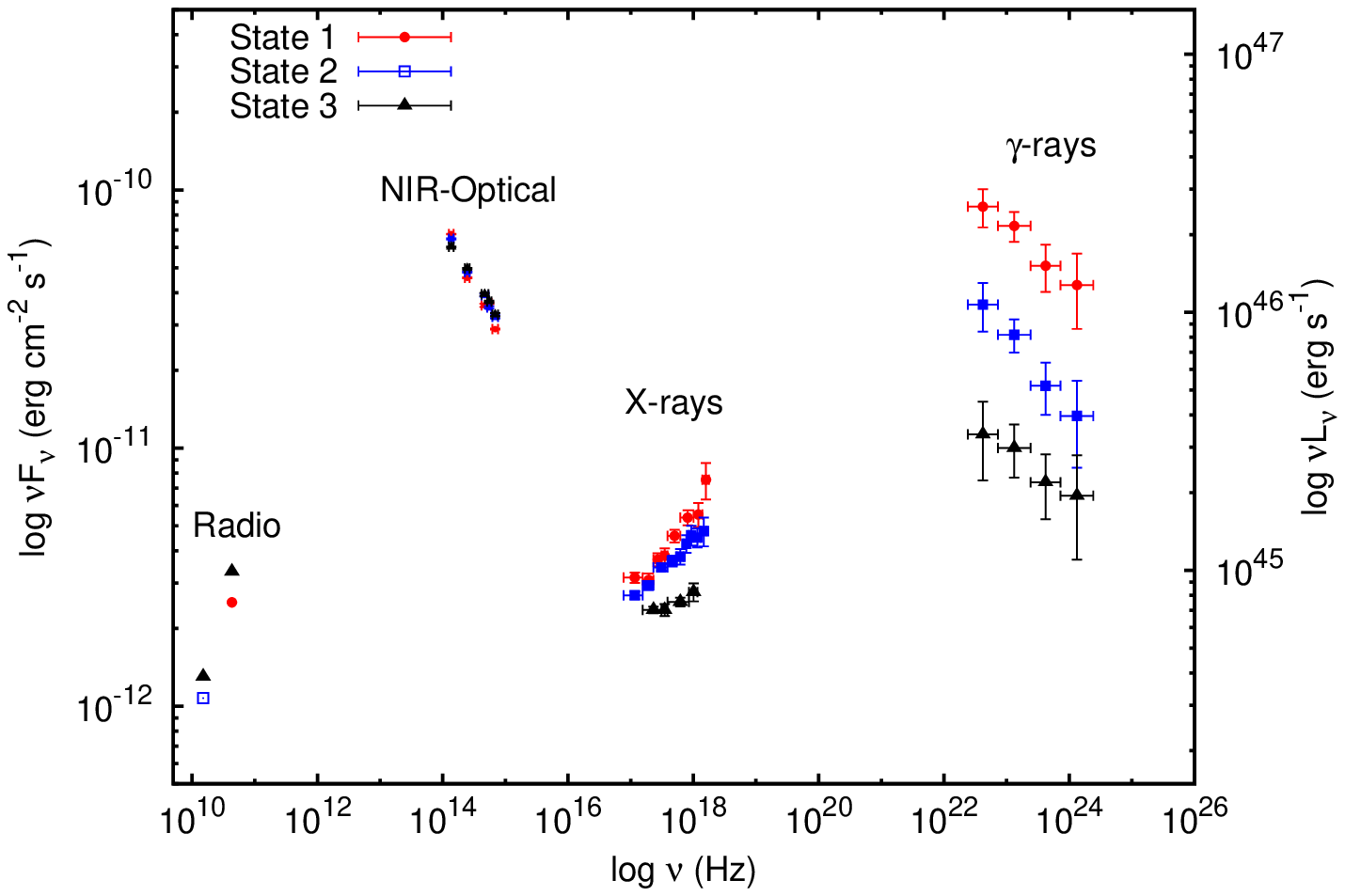}
\end{center}
\caption{Time averaged broadband SED of OJ 287 obtained for MJD: 55124-55131 (State 1), MJD: 55131-55152 (State 2) and MJD: 55152-55184 (State 3) 
(see Figure \ref{fig:lc}) during the 2009 $\gamma$-ray flare. State 1 SED corresponds to an average spectra from 20-27 October, 2009. The brightest $\gamma$-ray flare happened on October 22nd, 2009 while 
the XRT data for this state corresponds to October 25th, 2009. State 2 SED corresponds to an average spectra from October 28th-November 17th, 2009 while the State 3 SED corresponds to November 18th-
December 19th, 2009.}
\label{fig:multistatesed}
\end{figure}

\section{High Energy Emission Mechanism}\label{sec:emissmech}
To understand the X-ray and the $\gamma$-ray emission from OJ 287 during the flare, we adopt a simple model where
the emission region is assumed to be a sphere of radius $R$ moving down the jet at relativistic
speed ($\beta c$) with bulk Lorentz  factor $\Gamma=(1-\beta^2)^{-1/2}$ at 
an angle $\theta$ with respect to the line of sight of the observer. The emission
region is permeated with a tangled magnetic field $B'$ and populated by a broken power law distribution 
of particles described by (primed quantities are measured in the rest frame of the emission region)
\begin{equation}\label{eq:partdist}  
  N'(\gamma')d \gamma'=\left\{
  \begin{array}{l l}
    K \gamma'^{-p} d \gamma' & \quad ;\gamma'_{min}< \gamma'< \gamma'_b \\
    \kappa \gamma'^{-q} d \gamma' & \quad ;\gamma'_b< \gamma'< \gamma'_{max}\\
  \end{array} \right.
\end{equation}
with $\kappa=K\gamma_b^{'(q-p)}$. Here, $\gamma'_{min}m_ec^2$ and $\gamma'_{max}m_ec^2$ are the minimum and maximum energy 
of the particle distribution and $\gamma'_bm_ec^2$ is the break energy with $m_e$ being the rest mass of electron. 
The magnetic field and particle energy densities are related by
\begin{equation}\label{eq:eqpar}
U'_e = \eta U'_B
\end{equation}
where $U_e'$ is the particle energy density given by
\begin{equation}
U'_e = m_e c^2 \int_{\gamma'_{min}}^{\gamma'_{max}} \gamma' N'(\gamma')  d\gamma'
\end{equation}
and $U_B'$ is the magnetic field energy density
\begin{equation}
	U'_B =\frac{B_{eq}^{'2}}{8\pi}
\end{equation}
$\eta$ is a parameter and equipartition condition corresponds to $\eta\sim1$.
Particles lose their energy radiatively through synchrotron, SSC and/or EC processes. Due 
to relativistic motion and cosmological effects, the flux received by the 
observer on earth will be 
\begin{equation} \label{eq:obsflx}
 F(\nu) = \frac{\delta^3 (1+z)}{d_L^2}V' \epsilon'\left(\frac{1+z}{\delta} \nu \right)
\end{equation}
where $\delta = [\Gamma (1-\beta cos\theta)]^{-1}$ is the jet Doppler factor, $d_L$ is the luminosity
distance, $V'$ the volume of the emission region and $\epsilon'$ is the source emissivity due to 
different radiative processes.

Among the various parameters deciding the observed flux, the size of the emission region can be 
constrained through the variability timescale ($t_{var}$) as
\begin{equation}\label{eq:size}
  R' = \frac{\delta}{(1+z)} c t_{var}
\end{equation}
 For the present work, we consider the viewing angle of the jet for OJ 287 to be $\sim 3^\circ$ as estimated from VLBA
studies \citep{2005AJ....130.1418J} and the bulk Lorentz factor of the jet $\Gamma$ is chosen to be $12$ to obtain the observed  
superluminal velocity of 10.8c \citep{2011ApJ...726L..13A}\footnote{
However, it should be noted that our conclusion on emission mechanisms remain unchanged for the possible ranges of viewing angle
and bulk Lorentz factor inferred from the VLBA studies.}.
Under these assumptions and constraints, a plausible mechanism responsible
for the high energy emission can be argued based on the observed fluxes in optical, X-ray and
$\gamma$-ray energies. 

We have chosen the spectrum corresponding to the State 1 for present study and 
used the approximate analytical solution for synchrotron and EC emissivities \citep{2012MNRAS.419.1660S} to estimate
the source parameters.

\subsection{Synchrotron Self-Compton (SSC)} \label{sec:ssc}
In the interpretation based on SSC mechanism being operative, we consider the X-ray and $\gamma$-ray emission as resulting  from inverse
Compton scattering of synchrotron photons by the particle distribution described by equation
(\ref{eq:partdist}). The SSC peak ($\nu_{p,SSC}$) of the spectrum can then be related to the
synchrotron peak ($\nu_{p,syn}$) as
\begin{equation}\label{eq:sscpeak}
  \nu_{p,SSC} = \gamma_b^{'2} \nu_{p,syn}
\end{equation}
with 
\begin{equation}\label{eq:synpeak}
  \nu_{p,syn} = \frac{\delta}{1+z}\gamma_b^{'2} \nu_L
\end{equation}
Here $\nu_L = eB'/(2\pi mc)$ is the Larmor frequency.

The observed synchrotron flux for $\nu > \nu_{p,syn}$ 
can be approximated as 
\begin{equation}\label{eq:synflux}
  F_{syn}(\nu) \approx s(z,q) \delta^{(q+5)/2} B^{'(q+1)/2} R^{'3} \kappa \nu^{-(q-1)/2} Jy
\end{equation}
where $s(z,q)$ is a function of $z$ and $q$ (for $z=0.306$ and $q=3.54$, $s= 3.8\times10^{-47}$). 
Substituting equation (\ref{eq:size}) in (\ref{eq:synflux}) and choosing $p=2.42$ and $q=3.54$
(corresponding to photon indices $1.71\pm 0.05$ and $2.27\pm 0.10$), 
we can obtain the source magnetic field in terms of observed quantities as 
\begin{equation}\label{eq:magfld}
  B' \approx 0.08 \left(\frac{F_{5.5\times10^{14}Hz}}{6.6\times10^{-3} Jy}\right)^{0.44} 
		\left(\frac{\delta}{17.2}\right)^{-3.20} \left(\frac{t_{var}}{2.5 d}\right)^{-1.32}
		\left(\frac{\kappa}{2.4\times10^9}\right)^{-0.44}\left(\frac{\nu}{5.5\times10^{14}Hz}\right)^{0.56}
		 \; G
\end{equation}
The value of $\kappa$ is chosen to reproduce the SSC flux of $(5.5\pm0.5) \times10^{-11}\;Jy$ at $0.55\;GeV$.
Considering $\nu_{p,syn}\lesssim 10^{14}\;Hz$ (see Figure \ref{fig:multistatesed}) and using equations (\ref{eq:synpeak}) 
and (\ref{eq:magfld}) we get $\gamma'_b \lesssim 5.8\times10^3$. 
These estimated parameters correspond to an equipartition parameter $\eta\sim 215$ for assumed
$\gamma'_{min}=40$.

\begin{figure} 
\begin{center}
 \includegraphics[scale = 1]{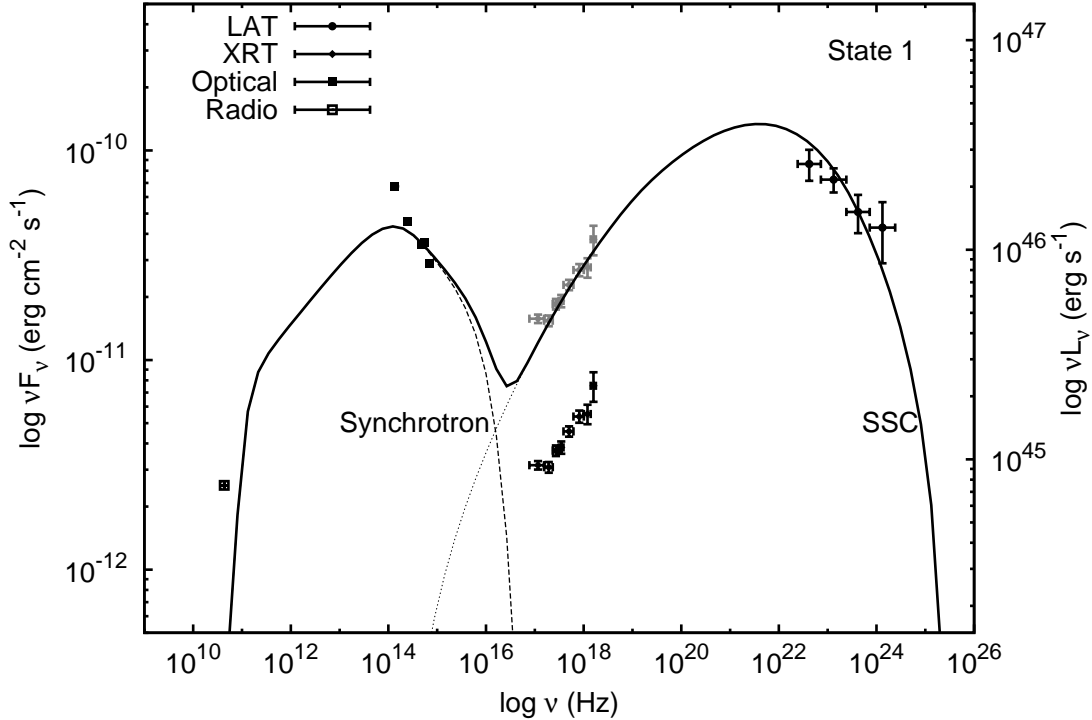}
\end{center}
\caption{Model spectrum due to synchrotron and SSC processes along with the SED corresponding to State 1 (see Figure \ref{fig:multistatesed}). 
The dashed and dotted curves represent the synchrotron and SSC components respectively while the solid curve is the total emission. The grey  
data points represent  five times the observed XRT flux (black point) during State 1 (see text).}
\label{fig:sscfit}
\end{figure}

 If we consider the SSC spectrum as a broken power law with indices $\alpha_X = 0.71 \pm 0.05$ 
and $\alpha_\gamma = 1.27 \pm 0.10 $, then the peak SSC frequency in SED can be obtained through X-ray and 
$\gamma$-ray fluxes as   
\begin{equation}\label{eq:comppeak}
  \nu_{p,SSC} = \left(\frac{F_{ssc}(\nu_\gamma)\nu_\gamma^{\alpha_\gamma}}{F_{ssc}(\nu_X)\nu_X^{\alpha_X}} \right)^{\left(\frac{1}{\alpha_\gamma-\alpha_X} \right)}
               \approx 3\times 10^{22} Hz
\end{equation}
From equation (\ref{eq:sscpeak}), this frequency corresponds to $\gamma'_b \gtrsim 1.5 \times 10^4$
 which contradicts our earlier condition on $\gamma'_b$. 
 However, considering that the X-ray observation was performed during the falling edge of the $\gamma$-ray flare, the X-ray flux 
of State 1 may be under predicted. If we increase the X-ray flux approximately five times, consistent with the factor of increase 
in the $\gamma$-ray flux corresponding to the highest and the lowest value, we can obtain $\gamma'_b \sim 4.1 \times10^3$. This satisfies
the $\gamma'_b$ constraint obtained earlier (using equations (\ref{eq:synpeak}) 
and (\ref{eq:magfld})). However, the parameters required to explain the SED deviate from the equipartition condition 
considerably\footnote{$\eta$ can at best be reduced to $\sim$30 by choosing superluminal velocity of 6.4c and a viewing angle
of $4.1^\circ$}.
In Figure \ref{fig:sscfit}, we plot the resultant spectrum due to synchrotron and SSC processes 
using the parameters described above.  For the model plot presented in Figure \ref{fig:sscfit} and 
the ones following (Figures \ref{fig:ecfit}, \ref{fig:allfit}, \ref{fig:allfit2} and \ref{fig:allfit3}) we have used the exact 
 description for radiative processes \citep{1995ApJ...446L..63D,1986rpa..book.....R} rather than the approximate 
 analytical expressions mentioned above and afterwards to analyse the different emission mechanisms.

\subsection{External Compton (EC)}

\begin{figure} 
\begin{center}
 \includegraphics[scale = 1]{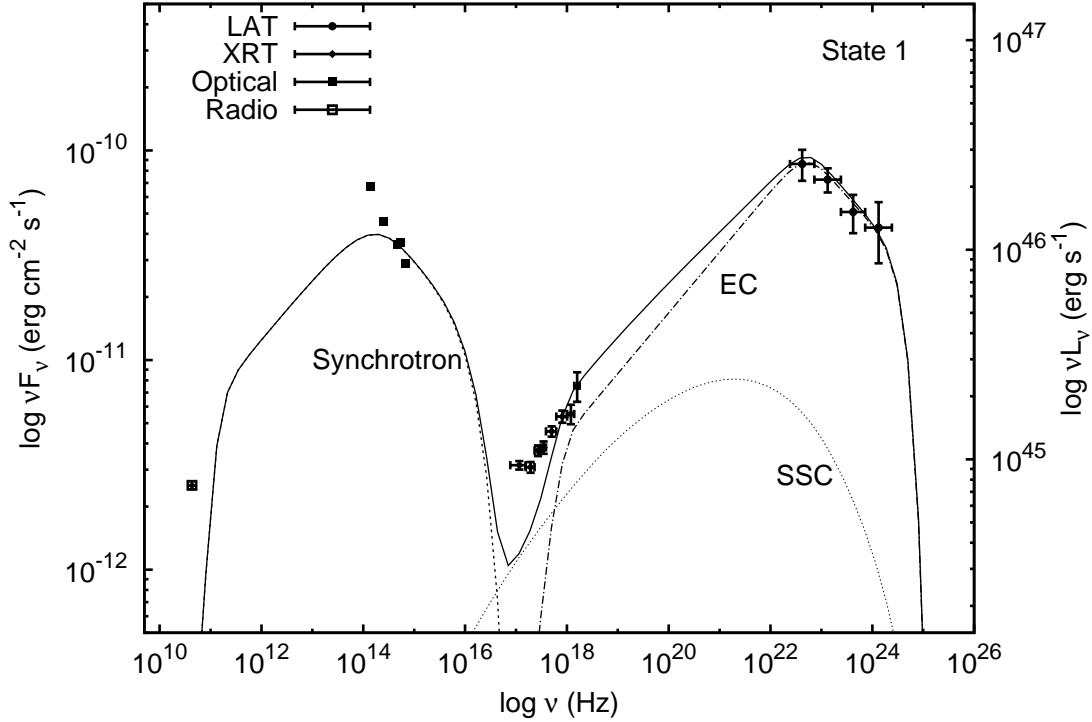}
\end{center}
\caption{Model spectrum due to synchrotron, SSC and EC processes along with the SED corresponding to State 1. The high energy emission is interpretated as 
a result of EC process only. As in Figure \ref{fig:sscfit}, the dashed and dotted curves represent synchrotron and SSC components respectively. The EC spectrum is represented 
by dash-dotted curve and the solid line is total spectrum due to all these emission processes.}
\label{fig:ecfit}
\end{figure}

In the EC scenario, the emission region moves through an external photon field and the high energy
emission is dominated by EC process rather than SSC process.
For simplicity we assume the external radiation to be of blackbody origin corresponding to 
a temperature $T_\ast$ (quantities with subscript $\ast$ are measured in the AGN frame). In the rest frame of emission region, 
the Lorentz boosted external photon field is scattered to high energy through inverse Compton process.
The peak of the EC spectrum due to the particle distribution given by equation (\ref{eq:partdist})
will be
\begin{equation}\label{eq:ecpeak}
  \nu_{p,EC} = \frac{\delta}{1+z}\gamma_b^{'2}(\Gamma \nu_\ast)
\end{equation}
where $\nu_\ast= 2.82 K_B T_\ast/h$ with $K_B$ and $h$ being Boltzmann and Planck constants.
The observed EC flux for $\nu>\nu_{p,EC}$ can be written as
\begin{equation}\label{eq:ecflux}
  F_{EC}(\nu)\approx c(z,q)\delta^{(q+5)/2} \Gamma^{(q+1)/2} R^{'3} \kappa \nu_\ast^{(q+5)/2}
	\nu^{-(q-1)/2} \; Jy
\end{equation}
where $c(z,q)$ is a function of $z$ and $q$ ($c(z,q)\sim8.3\times10^{-105}$ for $z=0.306$ and $q=3.54$).
Using equations (\ref{eq:synpeak}), (\ref{eq:synflux}), (\ref{eq:ecpeak}) and (\ref{eq:ecflux}) we
can obtain the magnetic field $B$ and $\nu_T$ in terms of observed quantities as
\begin{align}\label{eq:b_ec}
  B' \approx 0.3 \left(\frac{F_{0.55GeV}}{5.5\times10^{-11}Jy}\right)^{0.5}
	\left(\frac{F_{5.5\times10^{14} Hz}}{6.6\times10^{-3}Jy}\right)^{-0.5} 
	\left(\frac{\Gamma}{12}\right)
	\left(\frac{\nu_{p,syn}}{1.4 \times 10^{14} Hz}\right)^{2.14}
	\left(\frac{\nu_{p,ec}}{3\times10^{22} Hz}\right)^{-2.14} ~ G
\end{align}
Then from equation (\ref{eq:synpeak}) and (\ref{eq:ecpeak})
\begin{align} \label{eq:nu_ast}
  \nu_\ast \approx 1.5 \times10^{13} \left(\frac{F_{0.55GeV}}{5.5\times10^{-11}Jy}\right)^{0.5}
	\left(\frac{F_{5.5\times10^{14} Hz}}{6.6\times10^{-3}Jy}\right)^{-0.5}
	\left(\frac{\nu_{p,syn}}{1.4 \times 10^{14} Hz}\right)^{1.14}
	\left(\frac{\nu_{p,ec}}{3\times10^{22} Hz}\right)^{-1.14} ~ Hz
\end{align}

where $\nu_{p,EC}$ is obtained by considering the EC spectrum as a broken power law (refer equation (\ref{eq:comppeak})).
The lowest photon frequency of EC spectrum will then be 
\begin{align} \label{eq:nu_min}
  \nu_{min,EC} &=  \frac{\delta}{1+z}\gamma_{min}^{'2}(\Gamma \nu_\ast)\nonumber \\
	&\approx 3.4\times10^{17}\left(\frac{\delta}{17.2}\right)
	\left(\frac{\Gamma}{12}\right)
	\left(\frac{\gamma'_{min}}{12}\right)^2 
	\left(\frac{\nu_\ast}{1.5 \times10^{13}}\right)~Hz
\end{align}
However, this frequency is larger than the minimum observed frequency at X-ray energies ($1.2\times 10^{17}$) unless
one assume $\gamma'_{min}<\Gamma$ which is unphysical under shock acceleration 
theory \citep{2004MNRAS.349..336K,2002ApJ...564...97K}. Alternatively, $\nu_{min,EC}$ can be lowered
by reducing $\Gamma$ and $\delta$.
However, this demands an increase in $K$ to explain the observed EC flux. Since the SSC flux has a quadratic
dependence on $K$ \citep{2012MNRAS.419.1660S}, this will result in dominant SSC emission at 
X-ray energies and hence our EC interpretation fails.
Furthermore, the $\gamma'_b$ required to produce an EC peak frequency at $3\times10^{22} Hz$
by scattering of the soft photons at frequency  $\nu_\ast$ is $\approx 3.4\times10^3$. This again contradicts our constraint
obtained earlier (see \S \ref{sec:ssc}). Hence the interpretation of high energy emission by
EC process alone may not be a viable option though the deviation of the deduced quantities 
from the observed ones are marginal.  
The estimated value of $B'$ corresponds to an equipartition parameter $\eta = 2.3$ for $\gamma'_{min}=12$. 
The resultant spectrum due to synchrotron and dominant EC processes is shown in
Figure \ref{fig:ecfit}.

\subsection{SSC and EC Processes for High Energy Emission}

\begin{figure} 
\begin{center}
 \includegraphics[scale = 1]{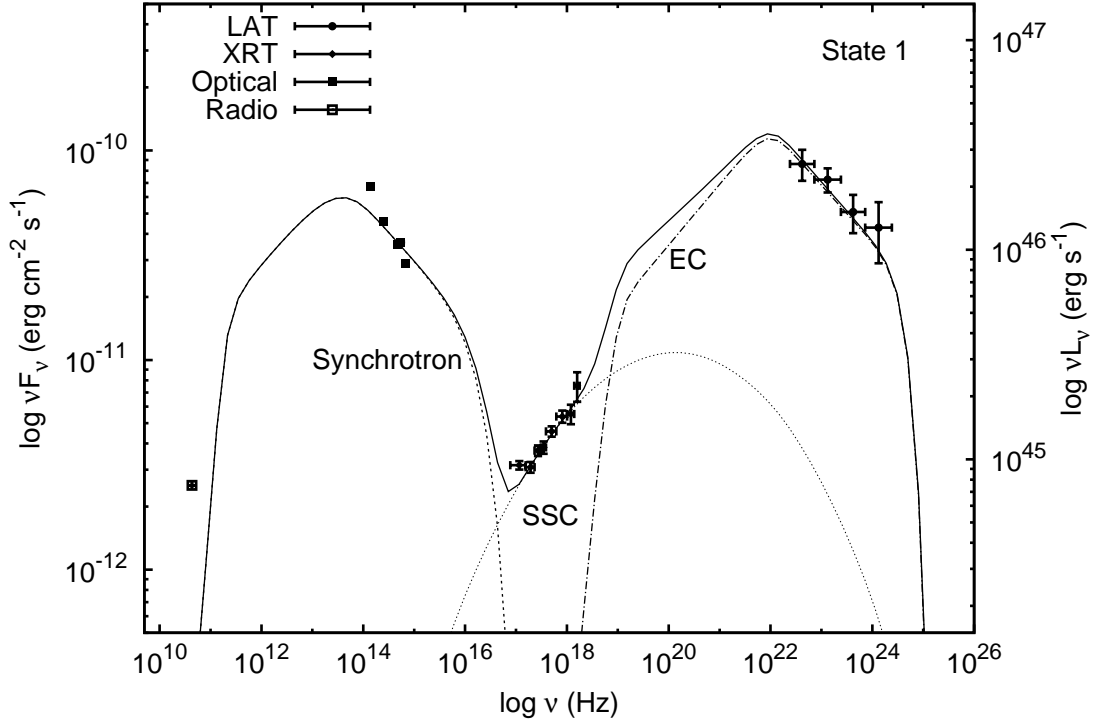}
\end{center}
\caption{ Model spectrum due to synchrotron, SSC and EC processes along with the SED of State 1. The high energy emission is interpreted as 
a result of both SSC and EC processes. The dashed, dotted and the dash-dotted curves represent the synchrotron, SSC and EC spectral components respectively. The solid curve
is the total emission from all the spectral components.}
\label{fig:allfit}
\end{figure}

\begin{figure} 
\begin{center}
 \includegraphics[scale = 1]{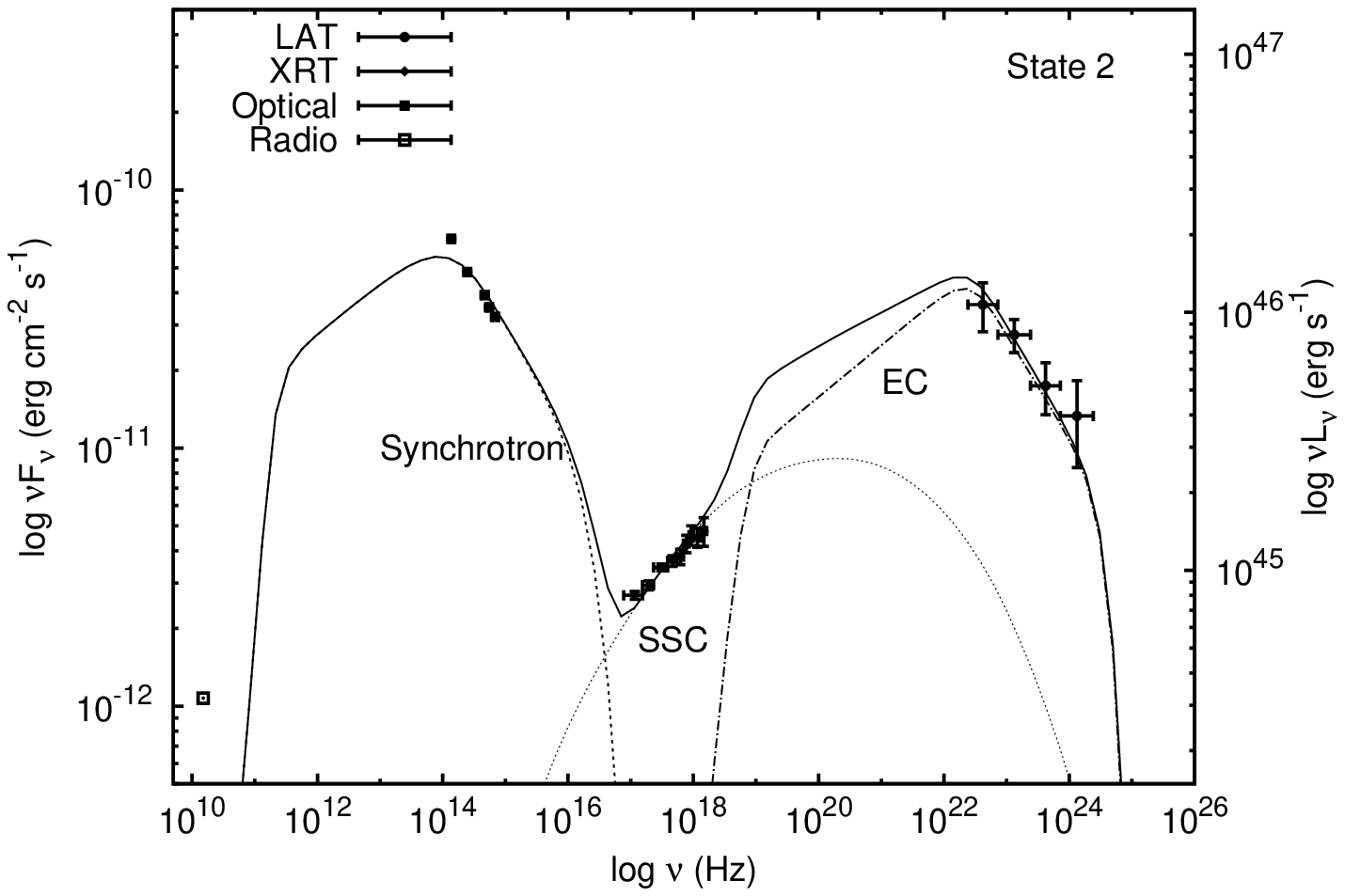}
\end{center}
\caption{ Model spectrum due to synchrotron, SSC and EC processes along with the SED of State 2. The high energy emission is interpreted as 
a result of both SSC and EC processes. The dashed, dotted and the dash-dotted curves represent the synchrotron, SSC and EC spectral components respectively. The solid curve
is the total emission from all the spectral components.}
\label{fig:allfit2}
\end{figure}

\begin{figure} 
\begin{center}
\includegraphics[scale = 1]{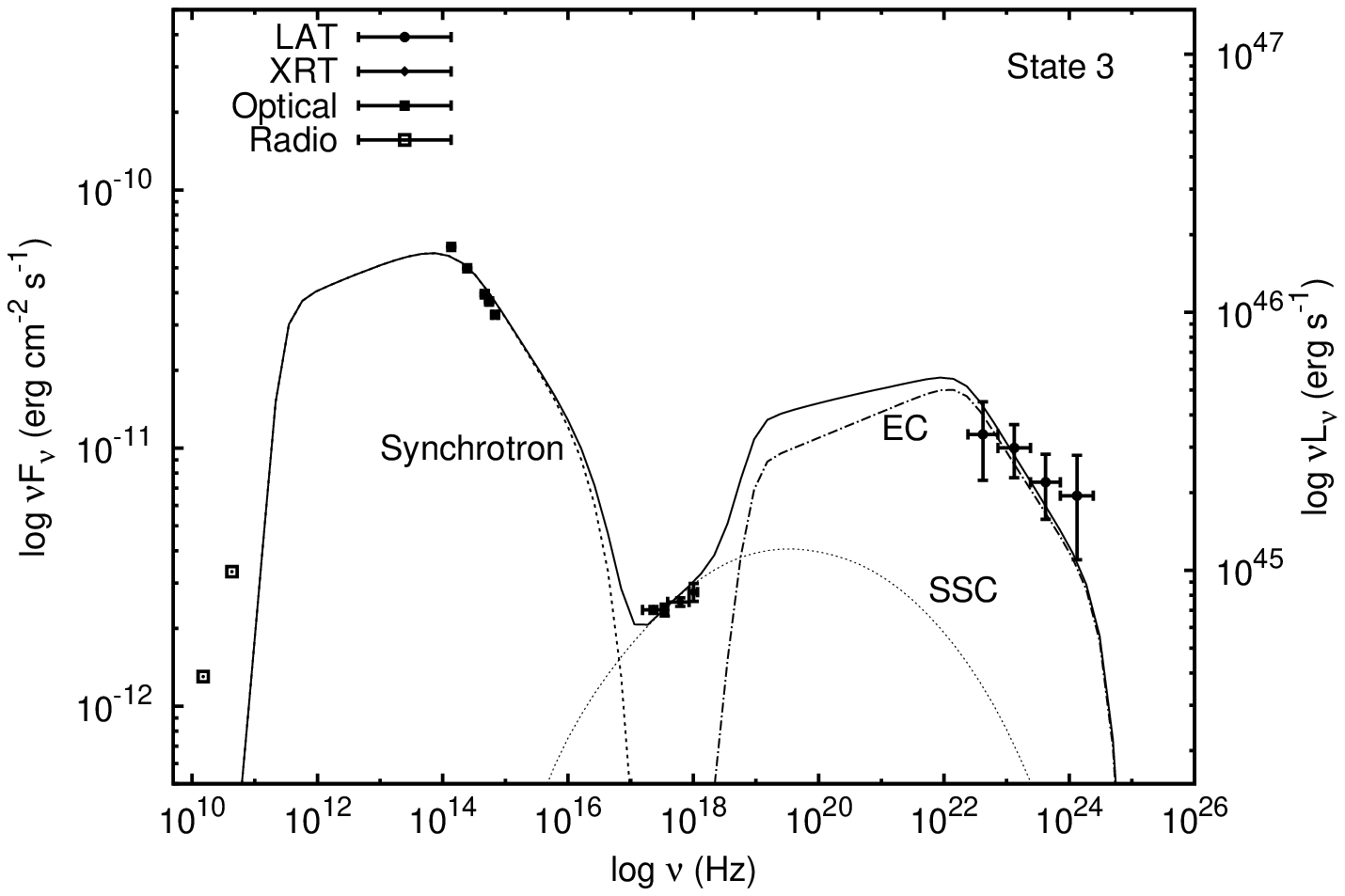}
\end{center}
\caption{ Model spectrum due to synchrotron, SSC and EC processes along with the SED of State 3. The high energy emission is interpreted as 
a result of both SSC and EC processes. The dashed, dotted and the dash-dotted curves represent the synchrotron, SSC and EC spectral components respectively. The solid curve
is the total emission from all the spectral components.}
\label{fig:allfit3}
\end{figure}

We consider the case where high energy emission is an outcome of both the SSC and EC
processes since individually either of these processes is unable to explain the observations satisfactorily.
Under this scenario, the X-ray emission is attributed to SSC process and the $\gamma$-ray 
emission to EC process. Then using equations (\ref{eq:eqpar}), (\ref{eq:synflux}) and (\ref{eq:ecflux})
for $\eta\sim1$ we obtain the temperature of the external photon field 
as\footnote{The obtained temperature can vary for different $\Gamma$ estimated from the allowed ranges of superluminal velocities and
viewing angles.} 
\begin{equation}
  T_\ast \approx 280 \left(\frac{F_{0.55GeV}}{5.5\times10^{-11}Jy}\right)^{0.23}
	\left(\frac{F_{5.5\times10^{14}Hz}}{6.6\times10^{-3} Jy}\right)^{-0.23} 
	\left(\frac{B'_{eq}}{0.4 G}\right)^{0.53}
	\left(\frac{\Gamma}{12}\right)^{-0.53}\, K
\end{equation}
The value of $B'$ is chosen to reproduce the SSC flux of $(9.1\pm 0.5)\times10^{-7}\;Jy$ at $2\;keV$.
The resultant spectrum of OJ 287 due to synchrotron, SSC and EC during State 1 is shown in the Figure \ref{fig:allfit}
along with the observed data. 
The physical parameters of the source governing the spectrum are given in Table \ref{tab:parameter}. 
A exercise similar to one described above for State 1 is repeated for State 2 and 3 and we have found that their high energy
spectra can be explained only if both SSC and EC processes are included. The resultant spectrum due to these emission processes 
are shown in Figures \ref{fig:allfit2} and \ref{fig:allfit3} and the corresponding parameters are given in rows 2 and 3 of Table \ref{tab:parameter}.
The spectrum of State 2 is reproduced using the equipartition parameter $\eta \sim 1$ whereas 
for State 3, during which the source was almost in quiescent state, we need to consider $\eta\sim 0.2$
to reproduce the observed spectrum.
Incidentally, we obtain almost similar temperature for the external photon field ($\sim 250 \,K$) in all the states.
The radio fluxes of all the states lie on synchrotron-self absorbed regime in  
the model plots (as is the case for most of the blazars) and the low energy break seen in synchrotron
spectrum is due to synchrotron-self absorption effect.

\begin{table}
\centering
 \begin{minipage}{140mm}
\caption{Model parameters and properties of the source} \label{tab:parameter}
\begin{tabular}{ l|cccccc|cc }
\hline 
  & \multicolumn{6}{|c|}{Parameters} \\ \cline{2-7}
State	& p	 & q & $\gamma_{b}'$ & $B'$ & $\eta$ & $T_\ast$ & $P_{jet}$ & $P_{rad}$\\
 \hline 
State 1	& 2.42 & 3.54	& $1.4\times10^3$ & 0.4 & 1.0 & 280 & $1.1\times10^{46}$ & $2.4\times10^{42}$ \\
State 2	& 2.60 & 3.80	& $2.6\times10^3$ & 0.4 & 1.0 & 224 & $1.3\times10^{46}$ & $2.3\times10^{42}$ \\
State 3 	& 2.80 & 3.76	& $2.2\times10^3$ & 0.7 & 0.2 & 235 & $1.1\times10^{46}$ & $2.4\times10^{42}$ \\
\hline
\end{tabular}
\\ 
Description of Columns: (1) Different states of the source (see Figure \ref{fig:lc}); (2) Particle spectral index before break (from X-ray data); (3) Particle spectral index after break (from LAT data);
(4) Break Lorentz factor of the electrons; (5) Magnetic field (in Gauss); (6) Equipartition factor (see \S 3); (7) Temperature of external photon field (in K);
(8) Total jet power ($erg/s$); (9) Total radiated power ($erg/s$). \\
Notes: For all the states we have chosen the emission region size  $R = 9\times10^{16}~cm$ (corresponds to a variability time of $\sim2.5$ days), 
bulk Lorentz factor $\Gamma = 12$, viewing angle $\theta =3^\circ$, minimum Lorentz factor of the electrons $\gamma_{min}'=40$  and maximum Lorentz factor of the electrons 
$\gamma_{max}'=3\times10^{4}$. Columns (2)-(7) are the parameters
governing the broadband spectrum of different states of the source whereas columns (9) and (10) are the jet and radiated power derived from these parameters.
 \end{minipage}
 \end{table}

\section{Discussion}\label{sec:discussion}
Our study suggests that the broadband spectra of OJ 287 observed during different stages of the $\gamma$-ray flare
in 2009 cannot be explained by considering the synchrotron and the SSC processes alone unless unlikely physical conditions are 
assumed. Hence an additional emission component is required to explain the high energy emission.
A plausible candidate for this additional component can be the EC scattering of soft photons 
external to jet. We assume this external photon field to be a blackbody radiation. With this
addition in the emission mechanisms we are able to reproduce the SED of OJ 287 
obtained during the different stages of the flare successfully. This result is similar to the
conclusion obtained through the empirical SED modelling of various LBL observed by \emph{Fermi-LAT}
\citep{2010ApJ...716...30A}. Models involving inverse Compton scattering of IR photons (EC/IR) 
from a dusty torus, proposed by the unified 
picture of the AGN \citep{1995PASP..107..803U}, are also used to explain the very high energy (VHE) 
emission from 3C 66A \citep{2011ApJ...726...43A} and ON 231 \citep{2010ApJ...716...30A}. 
The EC/IR interpretation is also proposed
for the BL Lac object AO 0235+164 since the SSC interpretation requires a very small covering factor of the
broad line regions (BLR) and IR dusty torus compared to the typical values of quasars \citep{2012ApJ...751..159A}. Earlier 
simultaneous observations of OJ 287 in X-ray and VHE during 2007 optical outbursts were modeled 
by \citet{2009PASJ...61.1011S}. No significant excess was reported at VHE during these observations and they explained
the broadband spectrum from radio to X-rays using synchrotron and SSC emission models. These observations were done before the 
launch of \emph{Fermi} and there were no instruments available to observe the source at MeV-GeV energies. 
However, inclusion of MeV-GeV flux, due to later observations by \emph{Fermi}, 
requires an additional emission component to explain the broadband SED (Figure 8 of \citet{2009PASJ...61.1011S}).

The physical parameters extracted by reproducing the observed spectrum of OJ 287 through synchrotron, SSC and EC
processes can be used to estimate the total
power of the jet. To do so we assume the jet is loaded with cold protons with their number density being equal 
to that of non thermal electrons. The  power of the jet can then be approximated as \citep{1997MNRAS.286..415C}
\begin{equation}
	P_{jet}=\pi R^{'2}\Gamma^2\beta c(U'_p+U'_B+U'_e)
\end{equation}
where $U'_p$ is the cold proton energy density. For the chosen set of parameters we find $P_{jet}\approx 10^{46} ~ergs/s$
which is approximately four orders of magnitude larger than the total power 
released as radiation $P_{rad}\approx10^{42}~ergs/s$ (Table \ref{tab:parameter}). Hence the
radiative processes are inefficient and most of the jet power can be carried to large scales. 
The X-ray jet of OJ 287 seen by \emph{Chandra} X-ray observatory has been studied by \citet{2011ApJ...729...26M}. Using parsec
scale viewing angle of $\sim 3.2 \,^\circ$, they derived the de-projected length to be greater than mega-parsec.
The X-ray emission from this mega-parsec scale jet is modelled as a result of IC/CMBR since at these length scales the dominant 
external photon field will be CMBR. The jet power estimated through their study is consistent with 
the one obtained above (Table \ref{tab:parameter}).

The \emph{Fermi} $\gamma$-ray spectrum obtained through the data analysis described in \S \ref{sec:obser} 
during the flare episode (State 1 and 2 combined) falls steeply beyond 10 GeV. In our model, the highest 
observed $\gamma$-ray photon energy is decided by $\gamma'_{max}$
provided the inverse Compton scattering happens in Thomson regime. In order to explain the X-ray spectrum 
as a result of inverse Compton emission, the synchrotron spectrum should fall before the X-ray energies (Figure \ref{fig:allfit}). 
This can constrain $\gamma'_{max}$ of the particle distribution which in our case is found to be $3\times 10^4$.
This high energy cut off in the particle spectrum at $\gamma'_{max}$ is reflected in the gamma ray spectrum  
at $\sim 10~GeV$ consistent with the observed \emph{Fermi} spectrum. This conclusion also states that 
the 2009 flare of OJ 287 is beyond the detectable 
threshold of ground based atmospheric Cherenkov telescopes operating at VHE. This result is consistent with the 
non-detection of OJ 287 by \emph{MAGIC}\footnote{Major Atmospheric Gamma-ray Imaging Cherenkov telescope} during 2007 
optical outbursts \citep{2009PASJ...61.1011S}.

Our analysis described in the previous section demands the presence of a warm
region at temperature $\sim 250 \,K$ around the emission region to 
explain the high energy emission from OJ 287. 
If we assume 
this region to be a spherical cloud surrounding the emission region then the 
extent of this region can be estimated from the flare timescale as
\begin{equation}
  R_{IR} \sim\frac{\Gamma^2c\;t_{var}}{1+z}\approx 0.23 \left(\frac{\Gamma}{12}\right)^2 \left(\frac{t_{var}}{2.5d}\right)\,pc
\end{equation}
The total IR luminosity of the cloud will then be
\begin{equation}
 L_{IR} = 4\pi R_{IR}^{2}\sigma_{SB}T_\ast^4 \approx 1.4 \times 10^{42} \,erg/s
\end{equation}
where $\sigma_{SB}$ is the Stefan-Boltzmann constant. This thermal IR luminosity is too small compared to
the continuum emission from the jet and hence the latter dominates the SED of OJ 287. 
Our result, therefore, is consistent with the understanding 
that the thermal IR emission is generally absent/weak in
BL Lac objects and their unification counterpart, 
FR I radio-galaxies \citep{1995PASP..107..803U, 2012ApJ...745L..27P, 1999A&A...349...77C}.
However, presence of a weak extended IR emission has been reported for the nearby FR I  radio-galaxies, Cen A \citep{2008ApJ...681..141R} and M87 \citep{2007ApJ...663..808P}.
Further, our estimated thermal IR luminosity is an order of magnitude smaller
than the IR upper limit obtained for the BL Lac object ON 231 \citep{ 2011ApJ...732..116M}. 
If we assume that the IR emitting cloud is powered by the radiation from an accretion disk having UV luminosity, $L_{UV}\sim 10^{46} ~erg/s$, then the
covering factor of the IR cloud as can be estimated as $L_{IR}/L_{UV}\approx 10^{-4}$. Using this covering factor 
we can obtain the location of the emission region from the central engine 
\begin{equation}
 D = 0.5 \left(\frac{L_{UV}}{L_{IR}} R_{IR}^{2}\right)^{1/2} 
	\approx 9 \left(\frac{L_{UV}}{10^{46}erg/s}\right)\left(\frac{T_\ast}{250 \, K}\right)^{-4}\,pc 
\end{equation}
This distance is comparable with the one obtained by \cite{2011ApJ...726L..13A} ($>14 ~pc$) through correlation 
study between 1 \emph{mm} radio and \emph{Fermi} $\gamma$-ray light curves and VLBA images.  Hence, our study suggests the presence
of a warm medium at temperature $\sim 250 \,K$ located at a distance $\approx 9 ~pc$
from the central engine of OJ 287. Previous studies of thermal emission from a dusty
environment of blazars and non-blazars also suggest the presence
of a hot dust at a temperature $\sim 800-1200 \, K$ extending up to a distance of $\lesssim 2 ~pc$ and 
a warm component at a temperature $\sim 150-300 \, K$ covering the hot region with the possible extension upto a
few tens of parsec \citep{ 2011ApJ...732..116M,2010MNRAS.408.1982L, 2004Natur.429...47J, 2009ApJ...705..298M}. 
A simulation study of this hot dust medium
employing two dimensional radiative transfer code and three dimensional radiative transfer code
using Monte Carlo technique also suggests a decrease in the temperature of the dust as one moves 
away from the central engine \citep{2005A&A...437..861S, 1992ApJ...401...99P}. Thus, it is possible
that the observation of OJ 287 presented in this paper probes the external regions of the dust emission.
A treatment similar to the one presented in this paper was used by
\citet{2012MNRAS.419.1660S} 
to conclude that the observed VHE emission from 3C 279 supports the EC/IR model. However, they obtained a temperature 
of the IR medium as
$\sim 900 \, K$ which is consistent with the hot dust at inner region of the torus.

The 2009 \emph{Fermi} $\gamma$-ray flare was also studied by \citet{2011MNRAS.412.1389N} 
who suggested the jet of OJ 287 to be associated with the lesser massive black hole of the SMBH binary system.
They used the fact that the observed variability timescale is much smaller than the light crossing time of black hole with a mass of $1.8\times10^{10}M_\odot$
but comparable to the one with a black hole mass of $1.3\times10^{8}M_\odot$. \citet{2011ApJ...726L..13A}, based on the luminosity 
ratio and simultaneity of optical and $\gamma$-ray flares concluded that the $\gamma$-ray emission is  
consistent with both the SSC and the EC/IR scenario. However, they favoured the SSC process since thermal IR emission is not 
detected from BL Lacs. Here we have studied the emission models in detail, estimating and constraining 
the governing parameters using various observational information, and  as already pointed out that both SSC as well as EC are required 
to interpret the high energy emission.

\section{CONCLUSIONS}
\label{sec:conc}
We have analyzed the archival X-ray and $\gamma$-ray observations of the BL Lac object OJ 287 during a $\gamma$-ray flare observed 
by \emph{Fermi} in 2009. Supplementing these data with the radio and near-IR-optical
data during the same period, we divided the multi-wavelength light curve into three parts: the flaring state, moderately 
active state and the quiescent state. The broadband SED corresponding to each state is then obtained and modeled using
synchrotron and inverse Compton emission processes. The main conclusions drawn from studies are:

1. The simple SSC interpretation of X-ray and $\gamma$-ray emission requires a broken power-law particle 
distribution with a large break energy. However, this is not supported by the synchrotron spectrum in the near-IR-optical energy-bands.

2. Interpretation of high energy emission based on EC process requires particles with Lorentz factor smaller than the bulk Lorentz factor of the jet to explain 
the lowest observed X-ray energy. However, this is not supported by the shock acceleration theory. Though the deviation of minimum
particle energy with the bulk flow encountered in this case is marginal, still the demand for the same cannot be achieved.

3. The high energy spectra, involving X-ray and $\gamma$-ray energies, can be readily explained by considering both SSC and EC 
processes together. Under this scenario the X-ray emission is attributed to the SSC process and the $\gamma$-ray emission to EC process.
To explain the $\gamma$-ray flux through EC process, we need the emission region to be buried inside a warm dusty region at a temperature 
of $\sim ~250 \,K$. If we consider the dusty environment of blazars to be illuminated by an accretion disk, then the location of the emission region 
should be $\sim ~ 9~ pc$ from the central engine. This distance is consistent with the constraints obtained from the millimeter-gamma
ray correlation studies and the VLBA maps of OJ 287.

The results presented in this work do not include the observational uncertainties. However, our conclusions on the emission processes remains unchanged
even if we deviate the observed fluxes and the other quantities within the allowed ranges.

PK thanks L. Resmi for suggestions and help on the analysis of LAT and XRT data.
This research has made use of data obtained from High Energy Astrophysics Science Archive Reasearch Center (HEASARC), maintained by NASA's 
Goddard Space Flight Center. Optical data from Steward Observatory, supported by Fermi Guest Investigator grants NNX08AW56G, NNX09AU10G, 
and NNX12AO93G along with IR-optical data from Yale \emph{Fermi}/SMARTS project were used. Radio data at 15 GHz from OVRO 40 M Telescope
funded in part by NASA and NSF and 43 GHz  data from BOSTON-VLBA gamma-ray blazar monitoring program funded by NASA are also used in this study.

\end{document}